\documentclass[conference]{IEEEtran}
\IEEEoverridecommandlockouts

\usepackage{bm}
\usepackage{amsfonts}
\usepackage{amsmath}
\usepackage{amssymb}
\usepackage{wasysym} 

\usepackage{multirow}
\usepackage{booktabs}
\usepackage{array}

\usepackage[noadjust]{cite}

\usepackage{graphicx}
\graphicspath{{./figure/}}

\usepackage{pgfplots}
\pgfplotsset{compat=1.15}

\usepackage[hidelinks]{hyperref}

\usepackage[linesnumbered, ruled, vlined, noend]{algorithm2e}

\SetCommentSty{mycommfont}
\SetKwIF{If}{ElseIf}{Else}{if}{}{else if}{else}{end if}%
\SetKw{Continue}{continue}
\SetKw{Break}{break}
\SetKw{Class}{Class}
\SetKw{And}{and}
\SetKw{Or}{or}
\SetKw{Elif}{elif}
\SetKw{True}{True}
\SetKw{False}{False}
\SetKw{To}{to}
\SetKw{Def}{def}
\SetKw{Not}{not}

\SetKwComment{Comment}{\color{green!50!black}// }{}

\newcommand{\FuncCall}[2]{\texttt{\bfseries #1(#2)}}
\SetKwProg{Function}{def}{}{}
\SetKwProg{Method}{def}{}{}

\newcommand{\algosize}{\small}  
\makeatletter
\renewcommand{\@algocf@capt@plain}{above}
\makeatother
\SetAlCapNameFnt{\small}
\SetAlCapFnt{\small}

\usepackage[caption=false]{subfig}
\newcommand{\V}{\mathcal{V}}
\newcommand{\E}{\mathcal{E}}

\newcommand{\PN}{\mathbb{PN}}
\newcommand{\CN}{\mathbb{CN}}
\newcommand{\pn}{\mathbb{P}}

\title{Spatio-Temporal Failure Propagation in Cyber-Physical Power Systems\\
    \thanks{This work was supported by NSF under Award Number 1808064.}
}

\date{\today}

\begin{document}

\author{
	\IEEEauthorblockN{Osman Boyaci}
	\IEEEauthorblockA{
		Electrical Engineering\\
		Texas A\&M University\\
		College Station, TX, 77843\\
		osman.boyaci@tamu.edu
	}  \and
	\IEEEauthorblockN{M. Rasoul Narimani}
	\IEEEauthorblockA{
		College of Engineering\\
		Arkansas State University\\
		Jonesboro, AR, 72404\\
		mnarimani@astate.edu
	} \and
	\IEEEauthorblockN{Katherine Davis}
	\IEEEauthorblockA{
		Electrical Engineering\\
		Texas A\&M University\\
		College Station, TX, 77843\\
		katedavis@tamu.edu
	} \and
	\IEEEauthorblockN{Erchin Serpedin}
	\IEEEauthorblockA{
		Electrical Engineering\\
		Texas A\&M University\\
		College Station, TX, 77843\\
		eserpedin@tamu.edu
	}
}

\maketitle

\begin{abstract}
Cascading failure in power systems is triggered by a small perturbation that leads to a sequence of failures spread through the system. The interconnection between different components in a power system causes failures to easily propagate across the system. The situation gets worse by considering the interconnection between cyber and physical layers in power systems. A plethora of work has studied the cascading failure in power systems to calculate its impact on the system.  Understanding how failures propagate into the system in time and space can help the system operator to take preventive actions and upgrade the system accordingly. Due to the nonlinearity of the power flow equation as well as the engineering constraints in the power system, it is essential to understand the spatio-temporal failure propagation in cyber-physical power systems (CPPS). This paper proposes an asynchronous algorithm for investigating failure propagation in CPPS. The physics of the power system is addressed by the full AC power flow equations. Various practical constraints including load shedding, load-generation balance, and island operation are considered to address practical constraints in power system operation. The propagation of various random initial attacks of different sizes is analyzed and visualized to elaborate on the applicability of the proposed approach. Our findings shed light on the cascading failure evolution in CPPS.
\end{abstract}


\section{Introduction}

The interaction between different components in cyber-physical power systems (CPPS) makes these systems more vulnerable to cascading failures as failure in one component can easily propagate into other components \cite{nature_cascading}. Electric Power Systems incorporate numerous communication devices, that enable full observability, controllability, and flexibility of power system operation. Furthermore, utilizing communication components in power systems increases the interaction surface between different components and paves the way for initiating cascading failures in CPPS. Cascading failure starts from a local failure and propagates into the system by overloading other components. For instance, the U.S. 2003 blackout was initiated to a large extent by the failure that initially occurred in the information and communications technology (ICT) system~\cite{Interconnected_CI}. Studying how the failures propagate in time and space is crucial for making power system more resilient. A plethora of research has dedicated to study the impact of cascading failures on the power system. However, the dynamics of cascading failures and how they propagate into the system is less studied. To predict and mitigate failure spreading in a CPPS, understanding their spatio-temporal propagation properties is crucial. To bridge this gap, we model a cascading failure propagation and analyze how cascading failures propagate into CPPS. To this end, we randomly attack different lines and force them to be out of service and then study the impact of these outages on the system. 

The mechanism of failure propagation in power systems strongly depends on the physics of the power systems, which is dictated by the nonconvex nonlinear power flow equations. Numerous studies have neglected the power flow equations in studying cascading failure propagation in power system~\cite{cascading_power_topology}. These studies focus on the topology of the system and leverage complex network theory to model failure propagation
~\cite{Cascading_failure_topology5,degree_distribution, structural_index}. Although methods based on the complex network theory require a much lower computational time, it is difficult to incorporate the physical mechanisms of cascading failure into the topological metrics.
Sandpile model has been employed to study the cascading failure characteristics in different types of complex networks~\cite{Sandpile1}. However, this model is not able to address the physics of power systems.  
In power grids, flows are driven by Kirchoff's laws, and cannot be described by a network flow model.
Accordingly, when a failure occurs in a power grid, the power flow is redistributed on the the rest of the network and some elements could overload and fail, leading to cascading failures. Thus, without detailed power grid information, the results of the methods based on complex network theory yield differ greatly and could result in misleading conclusions about the grid vulnerability~\cite{Cascading_failure_topology6}. Various studies consider load redistribution of a failed node among the in-service nodes, i.e., when a node fails,
the load it was carrying (right before the failure) is redistributed equally among the remaining nodes~\cite{Cascade_Failures_neighboring_redistribution}. This is not the cases in power grid, where the physics of the system is governed by the nonlinear power flow equations and thus a more sophisticated model needs to address the load redistribution after a contingency.  
There are numerous studies that have addressed the physics of power system in cascading failure problems via linearized power flow equations~\cite{soltan2, soltan3}.
Although DC power flow equations have been utilized to reasonably solve various problems in power systems but it has been shown that cascading failure simulations that underlie DC model (e.g., ignoring power losses, reactive power flow, and voltage magnitude variations) can lead to inaccurate and overly optimistic cascade predictions~\cite{soltan2}. More specifically, utilizing DC power flow to study cascading failure in large networks tends to overestimate the number of in-service components after stopping the failure propagation~\cite{soltan2} as DC power flow cannot properly capture nonlinear mechanisms like voltage collapse or
dynamic instability in power systems. Hence, using the DC model for cascade prediction may result in a misrepresentation of the gravity of a cascade~\cite{soltan2}. 
Multiple studies developed statistical models that use data from simulations~\cite{cascading_data_statistic_1,cascading_data_statistic_2} or historical cascades~\cite{cascading_data_statistic_3} to represent the general features of cascading. Statistical models are useful but cannot replace detailed simulations to fully understand cascading mechanisms.

To address these drawbacks, our method proposes an asynchronous algorithm that periodically propagates the failures between cyber and power layers. Furthermore, the full AC power flow equations along with the engineering constraints in power grids, e.g., load-generation balance, island operation criteria, etc. have been utilized to properly address the physics of power systems. Moreover, a realistic graph generation algorithm~\cite{boyaci2022generating} is employed to construct the underlying cyber graph in CPPS.  

\section{Propagation of Cascading Failure}

Before delving into the proposed approach for failure propagation, we need a realistic framework for cyber graph generation which is essential for modeling the failure propagation in CPPS.  
We recently proposed a realistic model for graph generation that mimics the features of a real-world communication system of a smart grid~\cite{boyaci2022generating}. The model leverages Hungarian algorithm~\cite{kuhn1955hungarian} to minimize the total cross edge distance in the graph which reduces the distance between corresponding cyber and power system nodes. We applied our approach in~\cite{boyaci2022generating} to generate realistic cyber graphs in this paper. 

To model the failure propagation in cyber-physical power system, we model the interconnected CPPS with a graph. In this connection, we first consider an undirected graph $\mathcal{G}(\mathcal{V},\mathcal{E})$ as a model of a power grid network. Here $\mathcal{V}$ is a
set of nodes, and $\mathcal{E}$ is a set of edges.
Similarly, the cyber layer is modeled with a graph with the same number of vertices that corresponding power system has, i.e, $\mathcal{V}$. The topology of the cyber graph, i.e., edge position, is determined by our proposed algorithm in~\cite{boyaci2022generating}. For sake of simplicity, we assume one connection between each node in the power systems with the corresponding node in the cyber layer. Note that the proposed model can be easily generalized to consider different typologies of the cyber graph, i.e., cyber graph with different number of nodes.

The proposed model is an asynchronous model in which faults propagate in only one subsystem, i.e., power or cyber graph, at a specific time step. Once the fault propagation is finished in one subsystem it can propagate into other subsystem through the linking edges that connect cyber and power graph nodes. The fault propagation in power and cyber layers acts differently as physics that governs these systems are totally different. For instance, in the cyber layer, after creation of islands, the largest island will be considered in-service and the other islands will be deleted from the system. Conversely, different islands in the power system can be operate simultaneously as long as they satisfy the power system constraints. Algorithm~\ref{alg:main} demonstrates the different phases of the proposed cascading failure approach where relevant method of the class ``'Grid' is called from Algorithm~\ref{alg:grid}. Phase \emph{A} of the proposed cascading failure approach models the failure propagation in power system after a random attack. Then phase \emph{B} reflects the impact of the topology changes in phase \emph{A} into the cyber layer by removing the cyber nodes that are connected to the deleted nodes in power system. Next, phase \emph{C} propagates the failure into cyber layer based on the node removals in Phase \emph{B} and update the topology of the cyber layer accordingly. 
\begin{algorithm}[h!]
	\SetAlgoNoLine
	\algosize
	\caption{Main of Cascading Failure}
	\label{alg:main}
	\SetKwInOut{Input}{Input}
	\SetKwInOut{Output}{Output}
	\SetKwFunction{FMain}{Main}
	\SetKwFunction{FGen}{Generate}
	\DontPrintSemicolon
	
	\Input{
		power network $\PN$,
		cyber network $\CN$,
		attacked buses $A_{\V}$,
		attacked branches $A_{\E}$
	}
	\Output{binary flag indicating the blackout}
	
	$grid \gets Grid(\PN, \CN)$ \tcp*{create the Grid obj.}
	grid.trigger\_failure\_in\_power($A_{\V}, A_{\E}$)\;
	\While{\True}{ 
		grid.prop\_failure\_in\_power() \tcp*{phase A}
		grid.prop\_failure\_to\_cyber() \tcp*{phase B}
		\lIf{$grid.stopped$} {\Break}
		grid.prop\_failure\_in\_cyber() \tcp*{phase C}
		grid.prop\_failure\_to\_power() \tcp*{phase D}
		\lIf{$grid.stopped$}{\Break}
	}
	$blackout \gets$ grid.check\_blackout() \;
	\Return $blackout$
\end{algorithm}

\vspace{-.2cm}

\begin{subequations} \label{OPF formulation}
\begin{align}
&\!\!\!\!\!\!\!\nonumber \text{subject to} \quad \left(\forall i\in\mathcal{N},\; \forall \left(l,m\right) \in\mathcal{L}\right) \\
\label{eq:pf1}
&\!\!\!\!\!\!\! P_i^g-P_i^d = g_{sh,i}\, V_i^2+\sum_{\substack{(l,m)\in \mathcal{L},\\\text{s.t.} \hspace{3pt} l=i}} \!P_{lm}+\!\!\sum_{\substack{(l,m)\in \mathcal{L},\\\text{s.t.} \hspace{3pt} m=i}} \!\!P_{ml}, \\
\label{eq:pf2}
&\!\!\!\!\!\!\! Q_i^g-Q_i^d = -b_{sh,i}\, V_i^2+\!\!\!\!\!\!\sum_{\substack{(l,m)\in \mathcal{L},\\ \text{s.t.} \hspace{3pt} l=i}} \!\!Q_{lm}+\!\!\!\sum_{\substack{(l,m)\in \mathcal{L},\\ \text{s.t.} \hspace{3pt} m=i}} \!\!\!\!Q_{ml},\\
\label{eq:pik}
&\!\!\!\!\!\!\! P_{lm} \!=\! g_{lm} V_l^2\! -\! g_{lm} V_l V_m\cos\left(\theta_{lm}\right)\! -\! b_{lm} V_l V_m\sin\left(\theta_{lm}\right),\\
\label{eq:qik}
&\!\!\!\!\!\!\! \nonumber Q_{lm} = -\left(b_{lm}+b_{c,lm}/2\right) V_l^2 + b_{lm} V_l V_m\cos\left(\theta_{lm}\right)\\ &\qquad\qquad  - g_{lm} V_l V_m\sin\left(\theta_{lm}\right),\\
\label{eq:pki}
&\!\!\!\!\!\!\!P_{ml}\! =\! g_{lm} V_m^2\! -\! g_{lm} V_l V_m\cos\left(\theta_{lm}\right)\! +\! b_{lm} V_l V_m\sin\left(\theta_{lm}\right),\\
\label{eq:qki}
&\!\!\!\!\!\!\!\nonumber Q_{ml} = -\left(b_{lm}+b_{c,lm}/2\right) V_m^2 + b_{lm} V_l V_m\cos\left(\theta_{lm}\right)\\ &\qquad\qquad  + g_{lm} V_l V_m\sin\left(\theta_{lm}\right).
\end{align}
\end{subequations}

Finally, phase \emph{D} reflects the impact of the topology changes in phase \emph{C} into the power layer by removing the power nodes that are connected to the deleted nodes in cyber layer. Phases \emph{A}-\emph{D} are continuously executed until no further removal happens in either the cyber or power layers. The cascading failure can be terminated if the number of removed nodes exceed a threshold.   

To make the proposed work more practical, the full AC power flow equations~\eqref{OPF formulation} are considered in this paper. Besides, other practical constraints including generation-load balance are considered for each island in the power system. Step-wise load shedding is also incorporated into the model to keep the islands that have not enough generation to satisfy their loads. This makes power system to not loose an island because of a small imbalance between generation and load and consequently withstand fault propagation for a longer period of time.

\noindent where $l$ and $m$ are both ends of each line $\left(l,m\right)\in\mathcal{L}$ in the system.
$g_{lm}+j b_{lm}$ and $j b_{c,lm}$ are the mutual admittance and shunt admittance in the  $\Pi$ model of power lines.
Define $\theta_{lm}=\theta_{l}-\theta_{m}$ for $(l,m)\in\mathcal{L}$.
The complex power flow into each line terminal $(l,m)\in\mathcal{L}$ is denoted by ${P}_{lm}+j{Q}_{lm}$.

\begin{algorithm}[h!]
	\SetAlgoNoLine
	\algosize
	\caption{Class Grid}
	\label{alg:grid}
	\SetKwInOut{Input}{Input}
	\SetKwInOut{Output}{Output}
	\SetKwFunction{FMain}{Main}
	\SetKwFunction{FGen}{Generate}
	\DontPrintSemicolon
	
	
	\tcp{Constructer method of the class}
	\Method{Grid::Grid($\PN, \CN$)}{
		$\PN \gets \PN$\;
		$\CN \gets \CN$\;
		$n, m \gets  |\PN.\V|, |\PN.\E|$ \tcp*{\# of components}
		$l \gets \langle\PN\rangle$      \tcp*{total active load in $\PN$}
		$\V_p^f, \E_p^f \gets \emptyset, \emptyset$ \tcp*{Failed power components}
		$\V_c^f, \E_c^f \gets \emptyset, \emptyset$ \tcp*{Failed cyber components}
		$stopped \gets \False$ \tcp*{Failure stopped}
	}
	
	
	\Method{Grid::trigger\_failure\_in\_power($A_{\V}, A_{\E}$)}{
		\ForEach{$v \in A_{\V}$}{
			mark bus $v$ in $\PN$ as out of service 
		}
		\ForEach{$e \in A_{\E}$}{
			mark branch $e$ in $\PN$ as out of service 
		}
	}
	
	
	\tcp{Phase A}
	\Method{Grid::prop\_power\_failure\_in\_power()}{	
		\While{\True}{
			\ForEach{island $\pn$ in $\PN$}{%
				\FuncCall{run\_power\_flow}{$\pn$}\;
				\FuncCall{remove\_overloaded\_branches}{$\pn$}\;
				\FuncCall{extract\_islands}{$\pn$}\;
			}
			\If{\textrm{there is no no new failure}}{
				\Break \;
			}
		}
		$\V_p^f, \E_p^f \gets$ determine failed components in $\PN$ \;
	}
	
	
	\tcp{Phase B}
	\Method{Grid::prop\_power\_failure\_to\_cyber()}{
		\If{$\V_p^f = \emptyset$}{
			$stopped \gets \True $
		}
		\Else{
			\ForEach{$u$ in $\V_p^f$}{%
				mark correspoding node $v$ as failed in $\CN$ \;
			}
		}
	}


	\tcp{Phase C}
	\Method{Grid::prop\_cyber\_failure\_in\_cyber()}{
		determine the connected components in $\CN$ \;
		keep only the giant component of $\CN$ \;
		mark other islands as failed in $\CN$ \;
		$\V_c^f, \E_c^f \gets$ determine failed components in $\CN$ \;
	}
	
	
	\tcp{Phase D}
	\Method{Grid::prop\_cyber\_failure\_to\_P($P$)}{
		\If{$\V_c^f = \emptyset$}{
			$stopped \gets \ \True $
		}
		\Else{
			\ForEach{$u$ in $\V_c^f$}{%
				mark correspoding node $v$ as failed in $\PN$ \;
			}
		}
	}
	
	
	\Method{Grid::check\_blackout($P$)}{
		\If{$|\PN.\V| < \frac{n}{2}$ \Or $|\PN.\E| < \frac{m}{2}$ \Or $\langle\PN\rangle < \frac{l}{2}$}{
			$\Return \ \True$\;
		}
		\Else{
			$\Return \ \False$\;
		}
	}
	
\end{algorithm}

The cascading failure initially triggered by the outage of certain nodes or lines in the power system (Alg.~\ref{alg:grid}, line 9-13).
These line outages change the network topology and might divide the power system network into separate components~\cite{MATCASC}.
If there is enough generation in each component to satisfy the load, then the component can operate.
If there is no generation in a separated component, it will be removed by force outage its nodes and branches. Likewise, if there was no load in a separated component, it will be removed by force outage its nodes and branches (Alg.~\ref{alg:util}, line 15-17).

For the components with both supply and demand, we first check the generation-demand balance.
If the total demand was greater than the generation, the demand would be reduced at all nodes by a common factor, i.e., load shedding.
Accordingly, if total generation was greater than the load, the generation would be reduced at all generation nodes by a common factor, i.e., generation curtailment (Alg.~\ref{alg:util}, line 22-23). 
We next check the line flows to determine the overflowed lines in the system. These lines will be removed from the system using a selected line outage rule (Alg.~\ref{alg:util}, line 8-11).  The cascade continues with the removal of those lines. If there is no new line failure in any of the remaining components, the cascade ends in the power system layer (Alg.~\ref{alg:grid}, line 20-21). 

\begin{algorithm}[h!]
	\SetAlgoNoLine
	\algosize
	\caption{Power utility functions}
	\label{alg:util}
	\SetKwInOut{Input}{Input}
	\SetKwInOut{Output}{Output}
	\SetKwFunction{FMain}{Main}
	\SetKwFunction{FGen}{Generate}
	\DontPrintSemicolon
	
	
	\Function{run\_power\_flow($\pn$)}{
		\ForEach{$s \in \{ 1.0, \ 0.95, \ \dots, \ 0.05\}$}{
			scale loads of $\pn$ with $s$ \;
			$result \gets$ \FuncCall{runpf}{$\pn$}\;
			\If{$result$}{
				mark island $\pn$ as succeded \;
			} 
		}
		mark island $\pn$ as failed \;
	}
	
	
	\Function{remove\_overloaded\_branches($\pn$)}{
		\ForEach{$e \in \pn.\E$}{
			\If{\textrm{$e$'s loading $>$ 100\%}}{
				mark $e$ as out of service \;
			} 
		}
	}
	
	
	\Function{extract\_islands($\pn$)}{	
		determine the connected components in $\pn$ \;
		\ForEach{$island \in \pn$}{
			\If{$\pn$ \textrm{has no generation or load}}{
				mark $\pn$'s component as out of service \;
				\Continue \;
			}
			\If{$\pn$ \textrm{has no slack bus}}{
				$g \gets$ highest generation capacity bus \;
				set $g$ as the slack bus \;
			}
		
			\If{\textrm{total generation} > \textrm{total load}}{
				apply generation curtailment\;
		    }
		}
	}
	
\end{algorithm}

\begin{figure}[h] 
	\centering
	\newcommand{\wdt}{0.47} 
	\subfloat[Triggered by buses \label{fig:blackout_v}]{
		\centering
		\includegraphics[width=\wdt\linewidth]{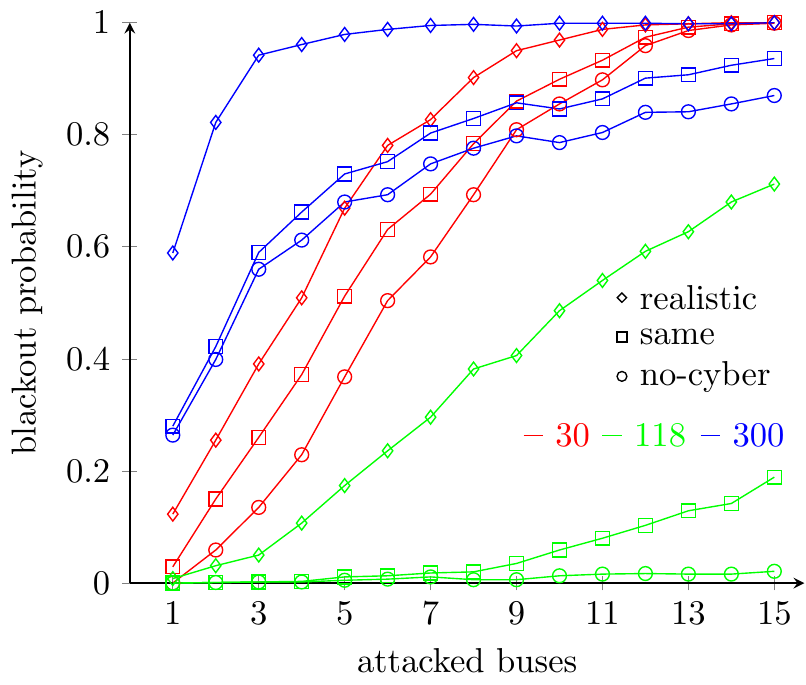}
	}
	\subfloat[Triggered by branches \label{fig:blackout_e}]{
		\centering
		\includegraphics[width=\wdt\linewidth]{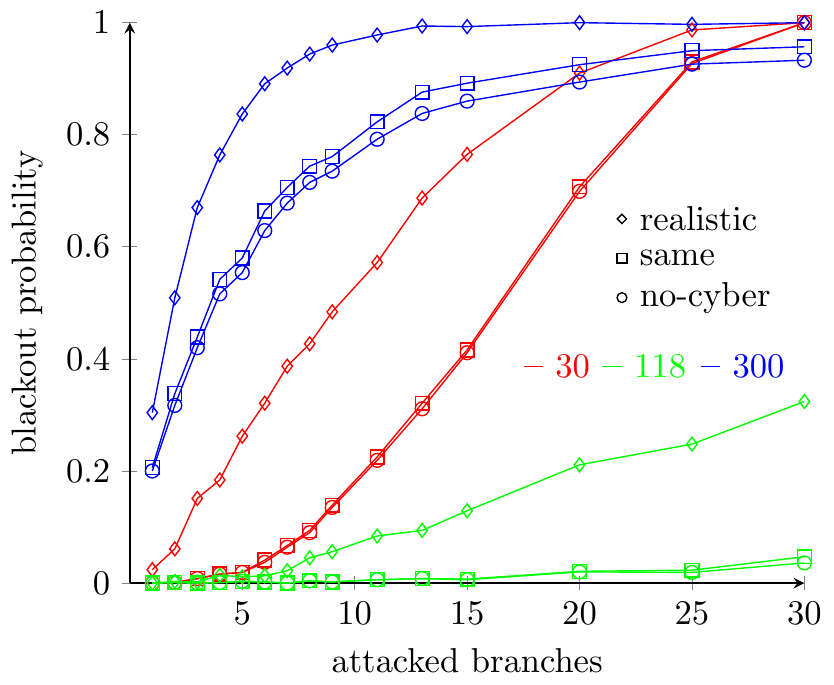}
	}
	\caption{Blackout probabilities of the cascading failure triggered by the number of initial attacked buses (a) and branches (b) for various systems and cyber networks. Each power network having 30- (red), 118- (green), and 300-bus (blue) simulated with three different cyber networks where there is no cyber graph (circle), cyber graph with the same topology as power system's graph (square), and cyber graph with a realistic topology (diamond), respectively. Blackout probabilities are calculated as the mean values of 1000 runs in which attacked buses or branches selected randomly.}
	\label{fig:blackout}
\end{figure}

\section{Results and Discussion}
This section demonstrates the effectiveness of the proposed approach using IEEE 30-, 118-, and 300-bus test cases.
All implementation was carried out in Python 3.8
on Intel i9-8950 HK CPU \@ 2.90GHz with NVIDIA GeForce RTX 2070 GPU.

Fig.~\ref{fig:blackout} demonstrates the blackout probability with respect to the number of the attacked nodes and branches for 30-, 118-, and 300-bus test cases. The blackout is defined as the lost of fifty percent of either the number of nodes, the number of branches, or total load in the system (Alg.~\ref{alg:grid}, line 40-44). It is clear that the probability of having blackout increases by the initial attack size as expected. There are three different plots in Fig.~\ref{fig:blackout} for each test case that are corresponding to the different topology of the underlying cyber graph of the power systems. The lines with circles, squares, and diamonds in Fig.~\ref{fig:blackout} represent the probability of occurring blackout in CPPS where there is no cyber graph, cyber graph with the same topology as the power system's graph, and cyber graph with a realistic topology, respectively. From Fig.~\ref{fig:blackout} it is clear that the topology of the underlying cyber graph in CPPS plays an important role in the dynamics of failure propagation.     

\begin{figure*}[h!] 
	\centering
	\newcommand{\wdt}{0.42} 	
	\newcommand{\n}{118} 		
	\newcommand{\vsp}{0.18cm} 	
	\subfloat[initial system]{
		\centering
		\includegraphics[width=\wdt\linewidth]{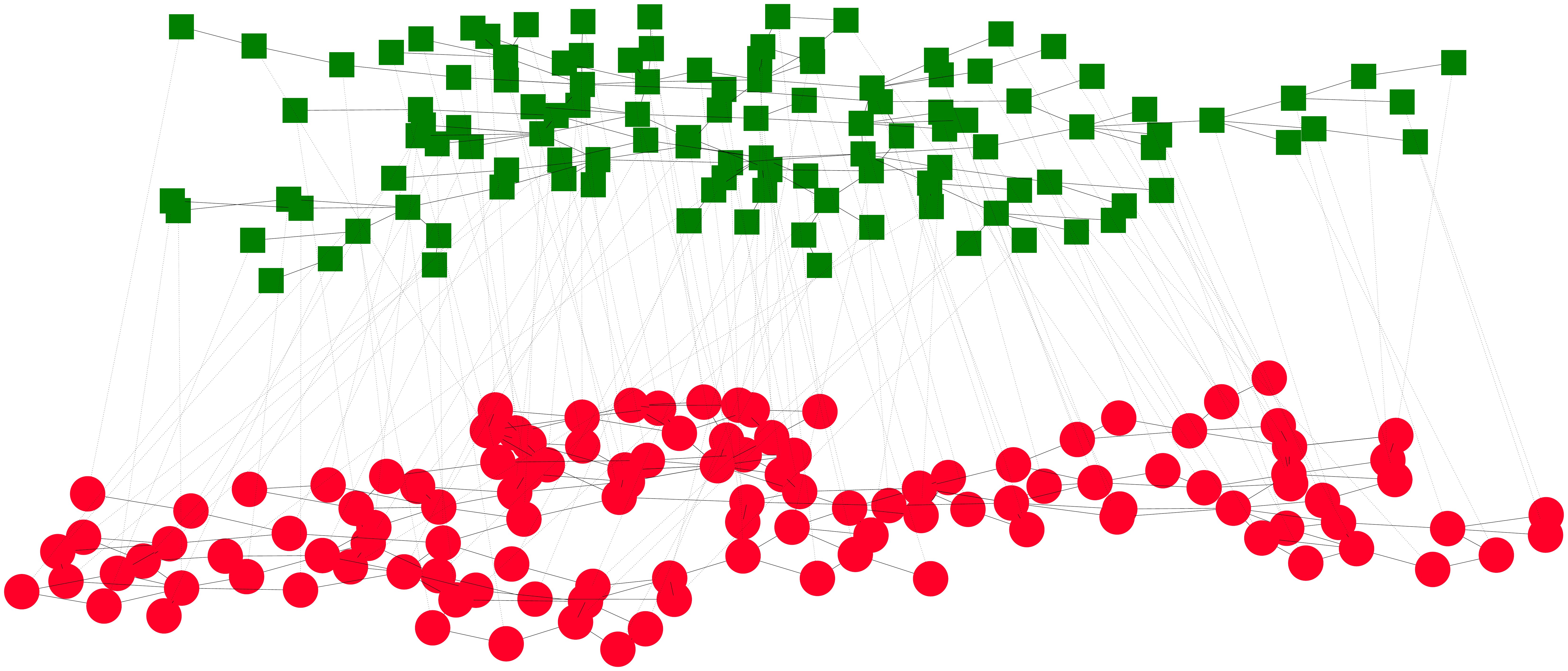}
	}
	\subfloat[failure triggered]{
		\centering
		\includegraphics[width=\wdt\linewidth]{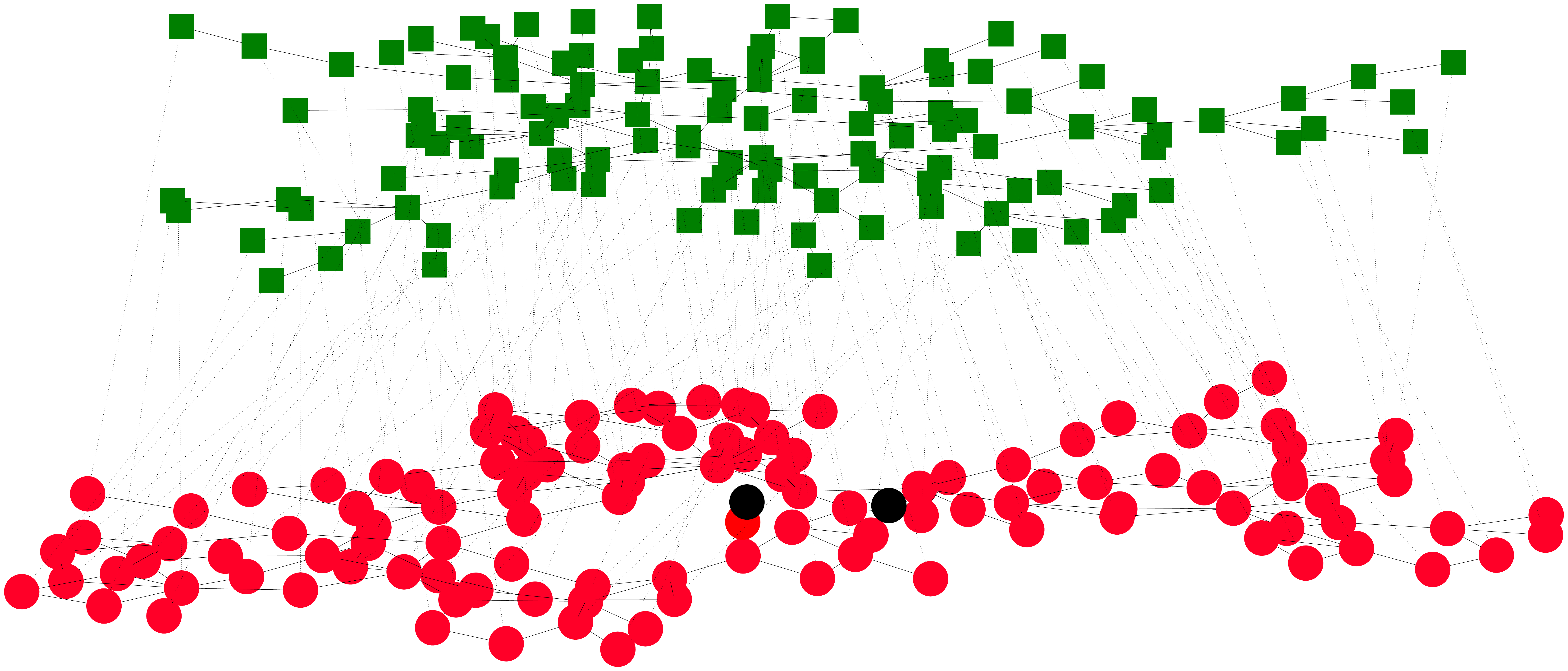}
	} \\ \vspace{\vsp}
	\subfloat[phase 1-A]{
		\centering
		\includegraphics[width=\wdt\linewidth]{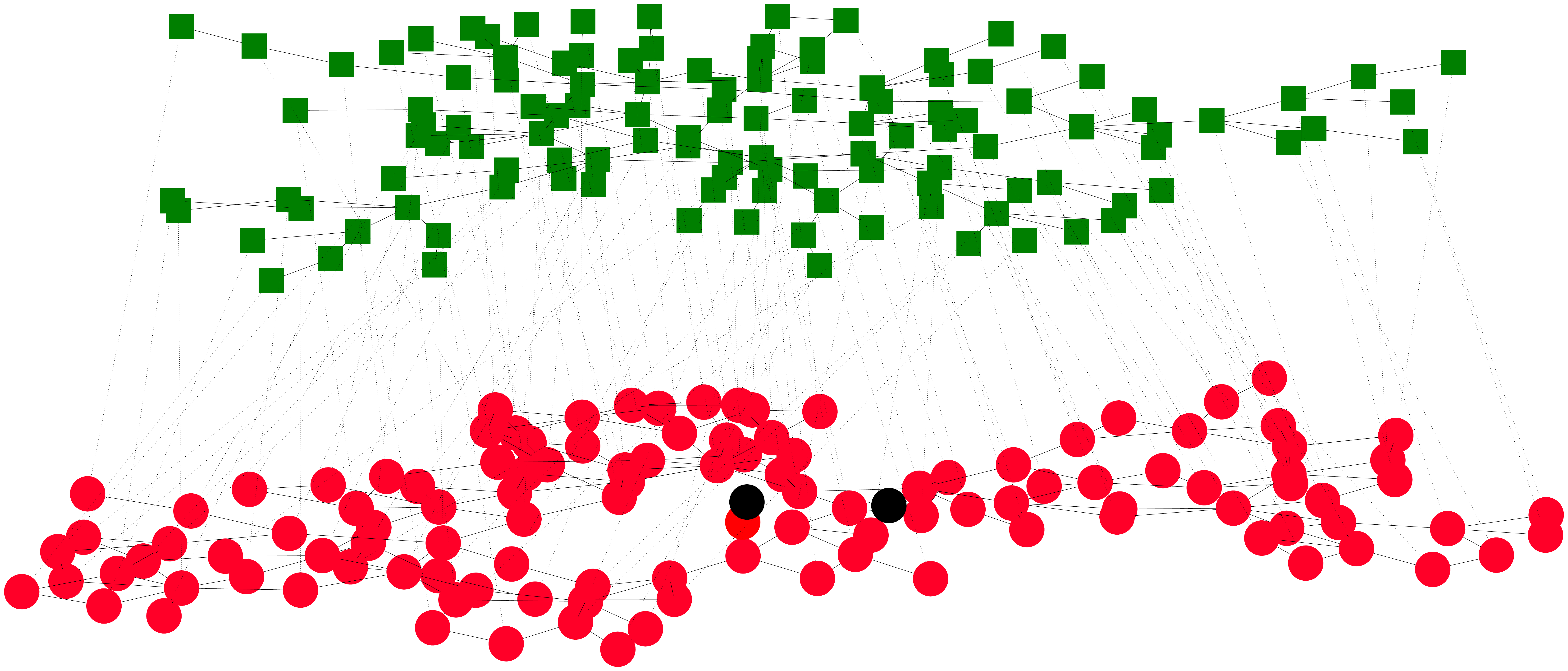}
	}
	\subfloat[phase 1-C]{
		\centering
		\includegraphics[width=\wdt\linewidth]{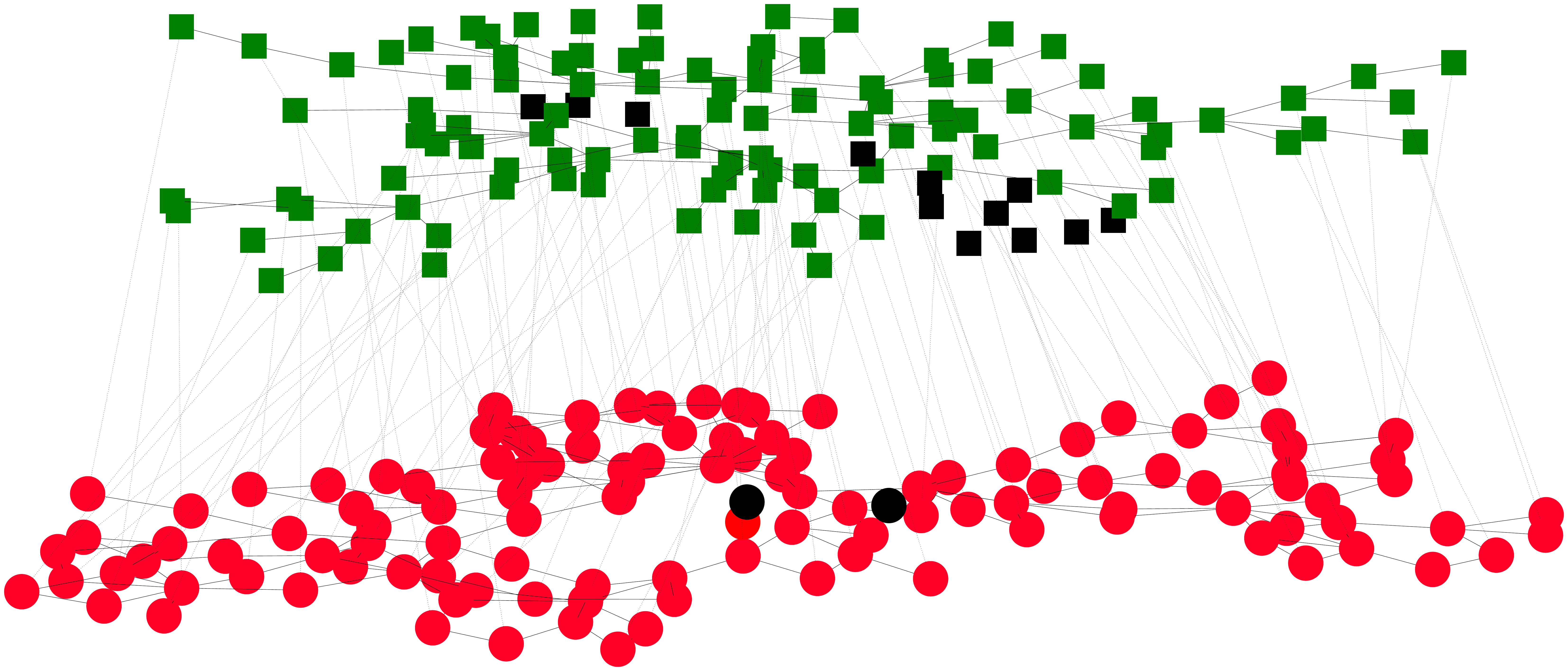}
	} \\ \vspace{\vsp}
	\subfloat[phase 2-A]{
		\centering
		\includegraphics[width=\wdt\linewidth]{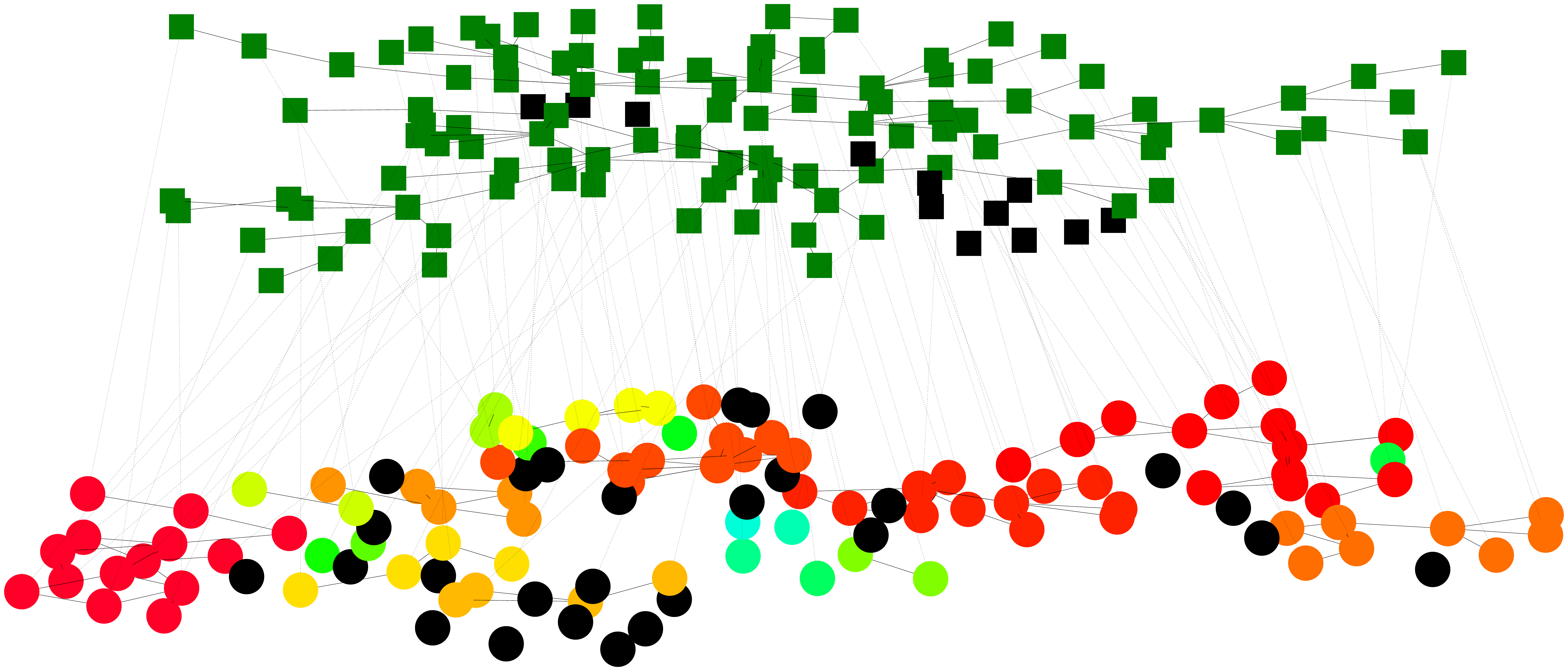}
	}
	\subfloat[phase 2-C]{
		\centering
		\includegraphics[width=\wdt\linewidth]{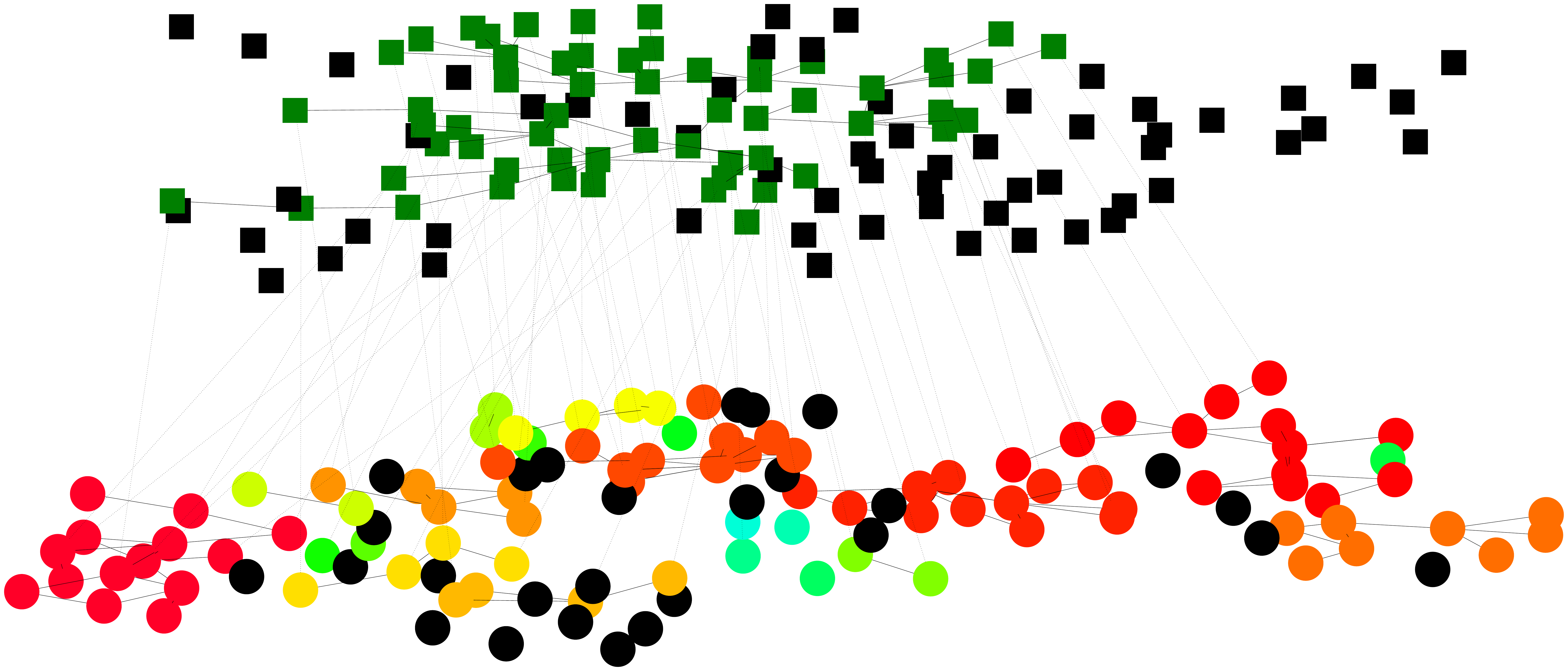}
	}\\ \vspace{\vsp}
	\subfloat[phase 3-A]{
		\centering
		\includegraphics[width=\wdt\linewidth]{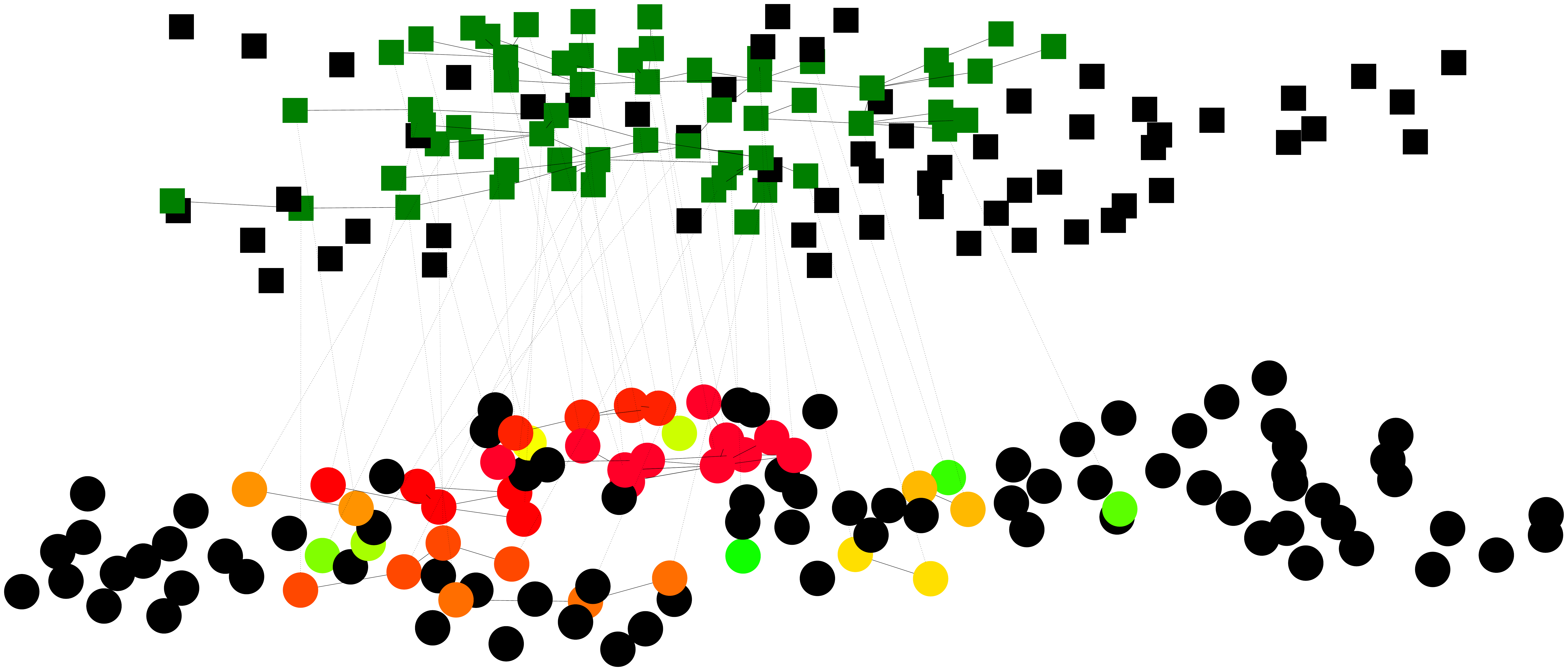}
	}
	\subfloat[phase 3-C]{
		\centering
		\includegraphics[width=\wdt\linewidth]{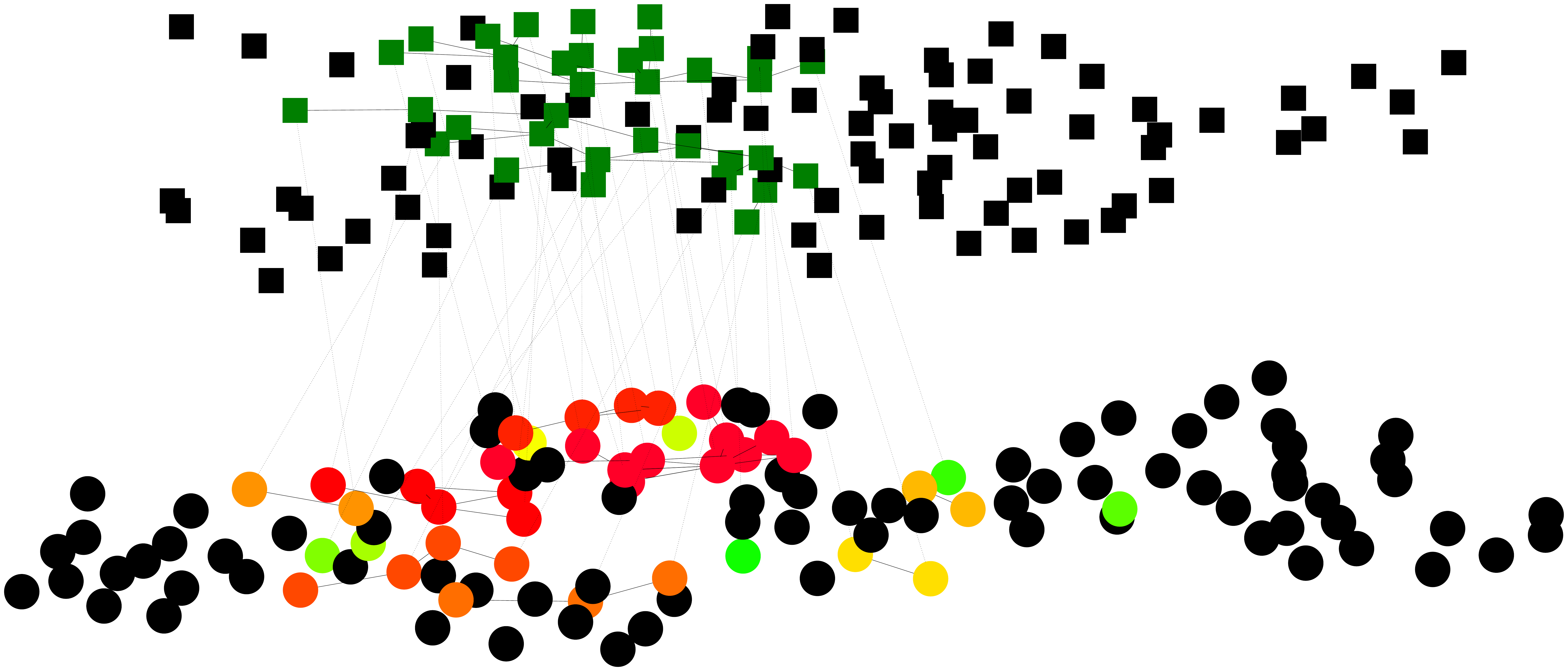}
	}\\ \vspace{\vsp}
	\subfloat[phase 4-A]{
		\centering
		\includegraphics[width=\wdt\linewidth]{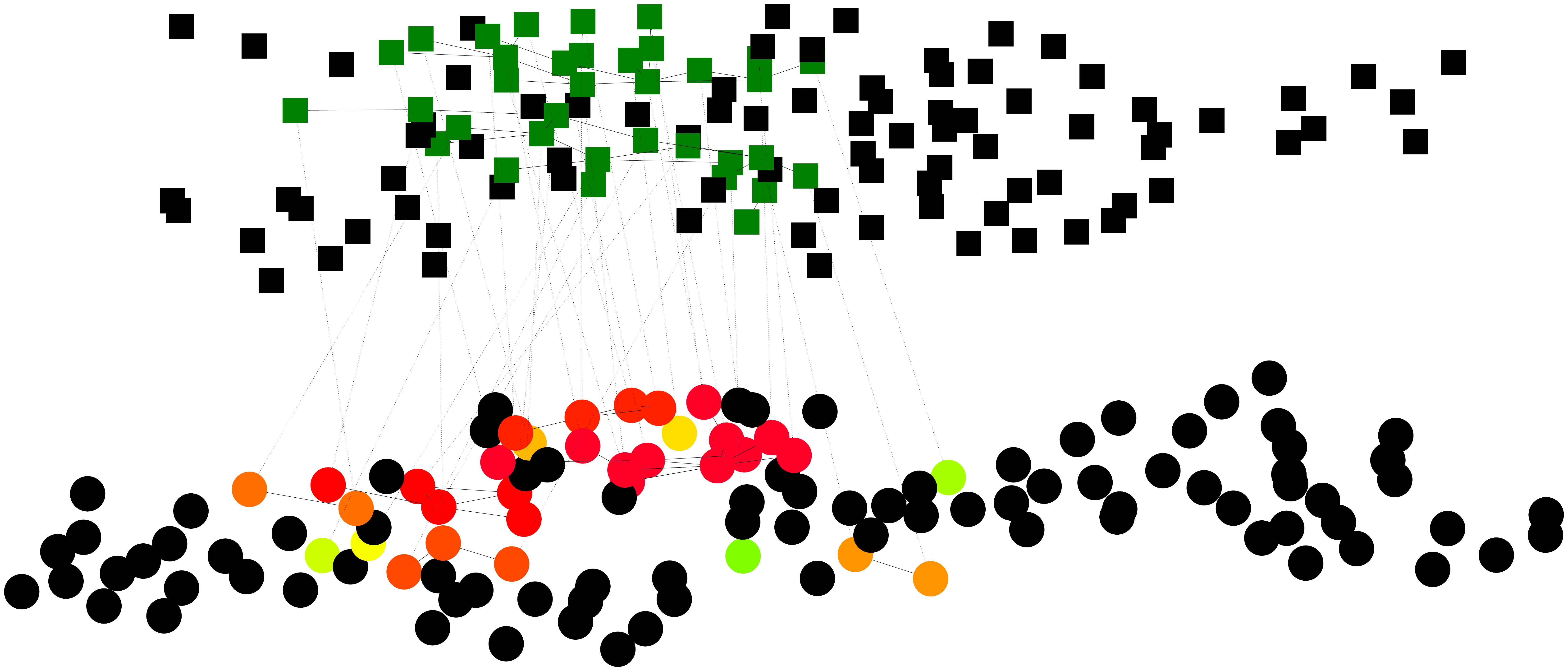}
	}
	\subfloat[phase 4-C]{
		\centering
		\includegraphics[width=\wdt\linewidth]{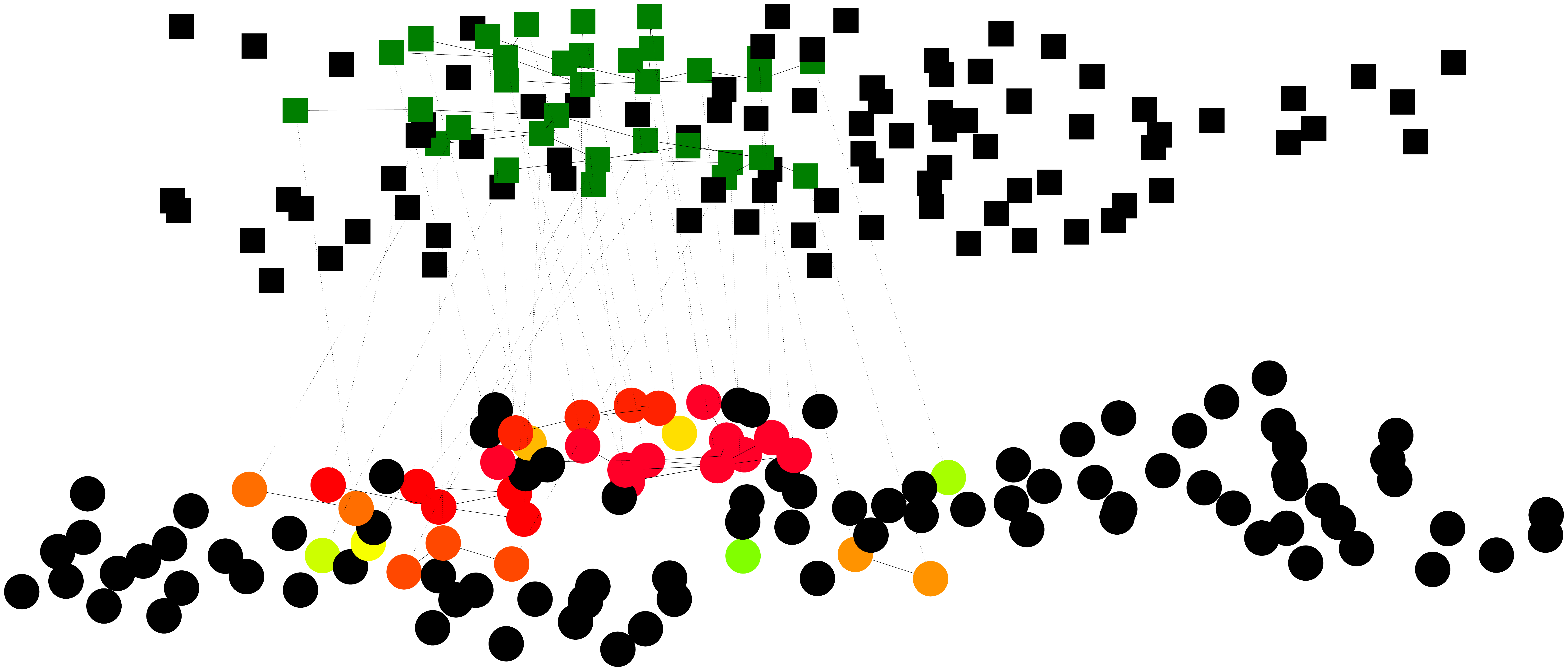}
	}\\
	\caption{Iterations of a cascading failure for the IEEE 118-bus test system where the power and cyber nodes are depicted with circles at the bottom layer and square at the top layer, respectively. Different islands are plotted with different colors and failed nodes are always given in black. Note that failure triggered in two power buses (b) and eventually stopped after four iterations (j). Due to the space limitation, only phases A and C  are depicted.}
	\label{fig:it118}
\end{figure*}

\begin{figure*}[h!] 
	\centering
	\newcommand{\wdt}{0.37} 	
	\newcommand{\n}{300} 		
	\newcommand{\vsp}{0.09cm} 	
	\subfloat[initial]{
		\centering
		\includegraphics[width=\wdt\linewidth]{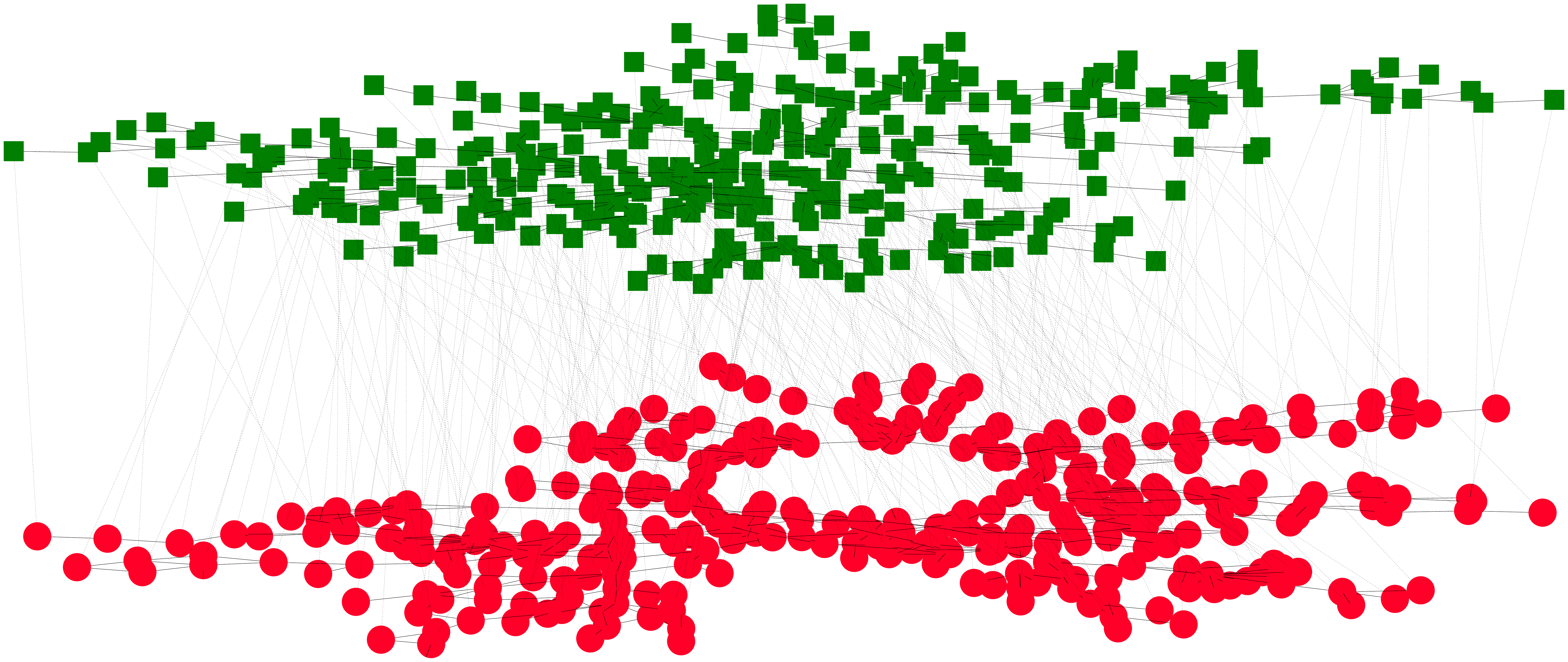}
	}
	\subfloat[triggered]{
		\centering
		\includegraphics[width=\wdt\linewidth]{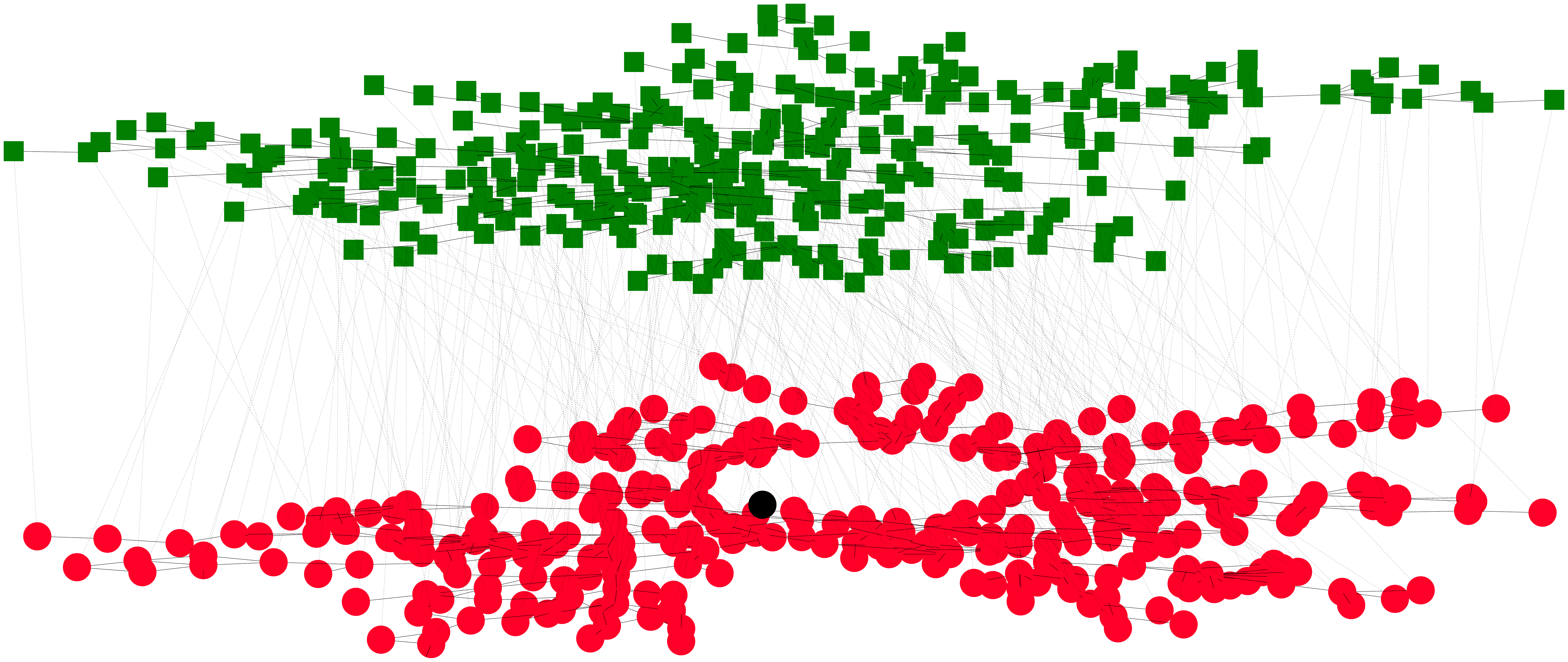}
	}
	\\ \vspace{\vsp}
	\subfloat[1A]{
		\centering
		\includegraphics[width=\wdt\linewidth]{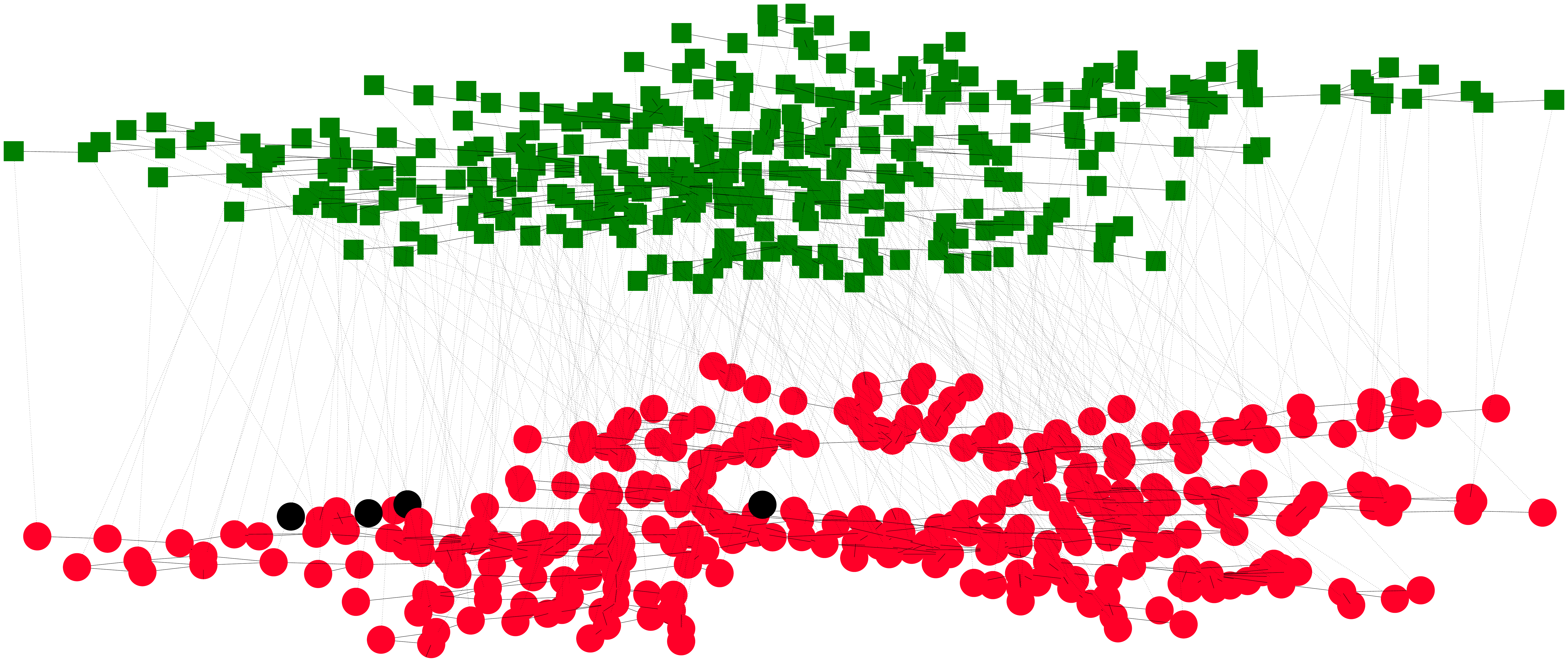}
	} 
	\subfloat[1C]{
		\centering
		\includegraphics[width=\wdt\linewidth]{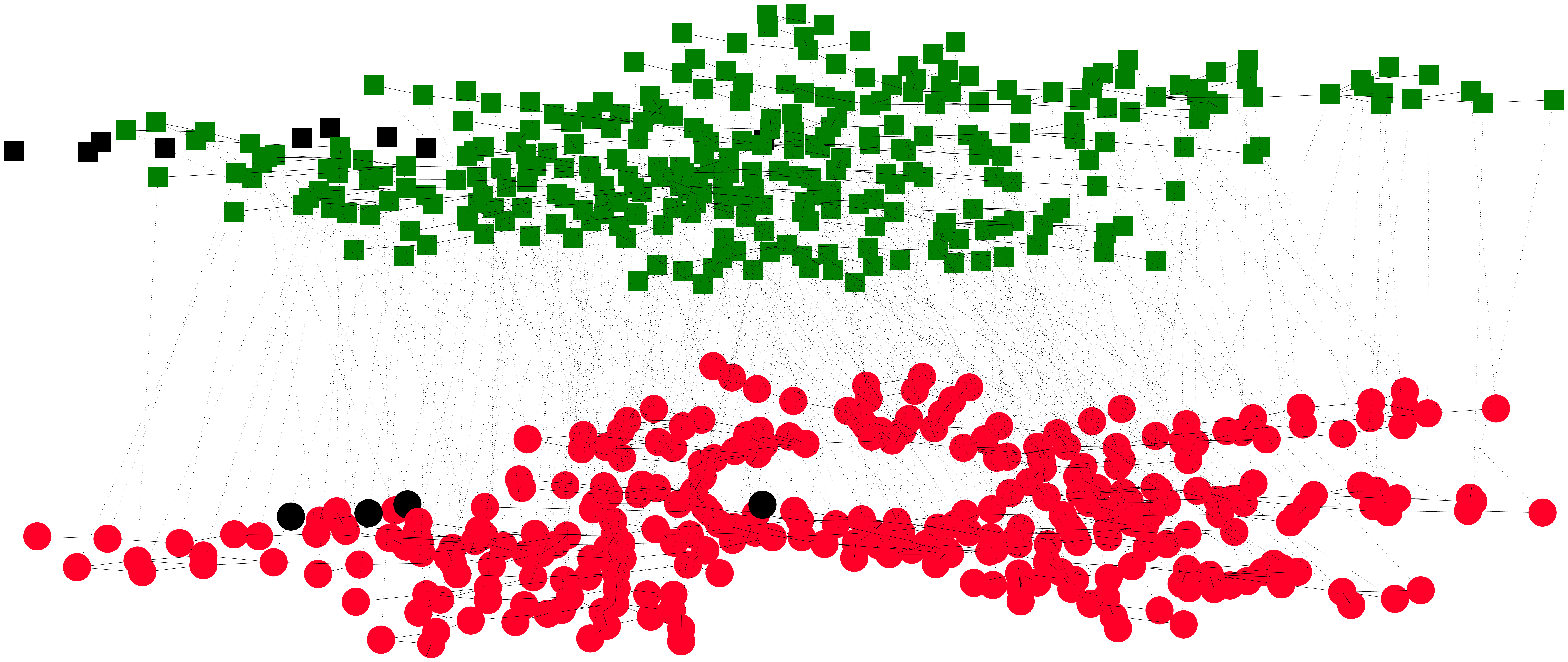}
	}
	\\ \vspace{\vsp}
	\subfloat[2A]{
		\centering
		\includegraphics[width=\wdt\linewidth]{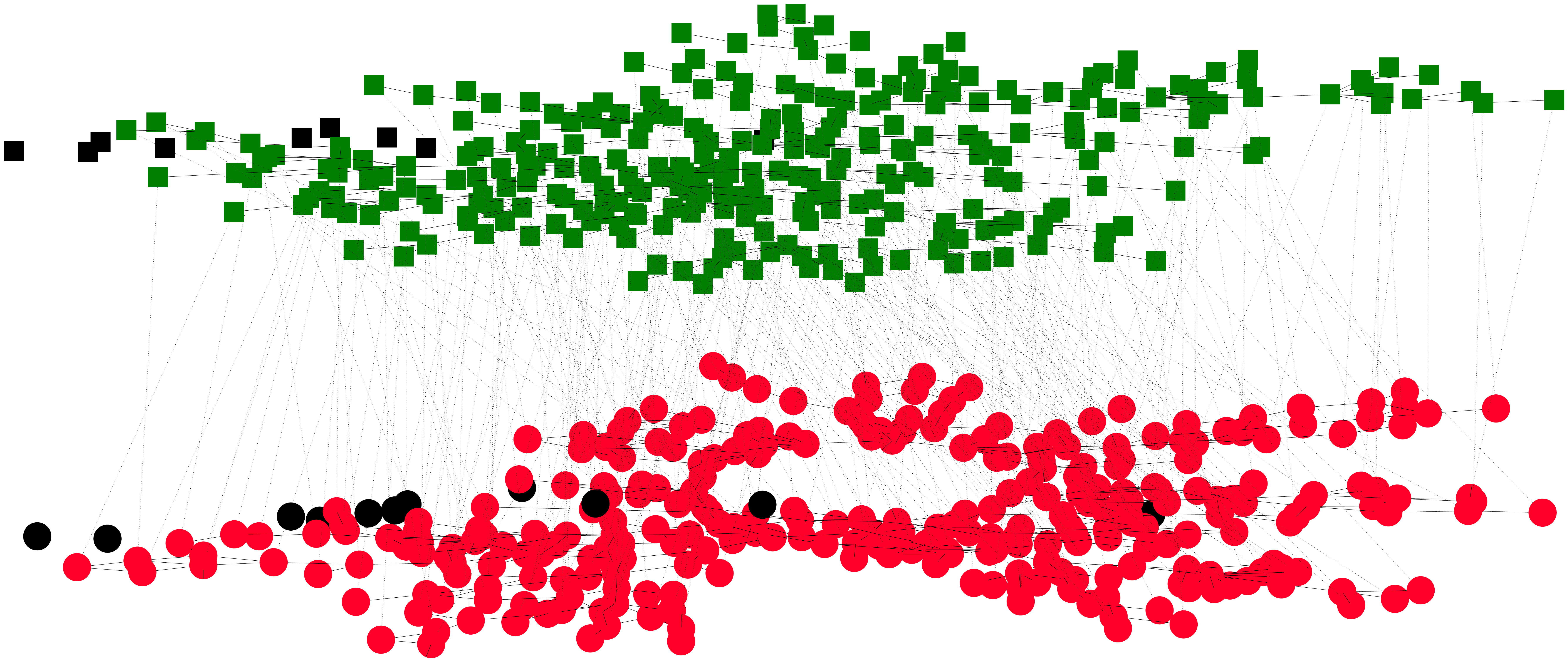}
	}
	\subfloat[2C]{
		\centering
		\includegraphics[width=\wdt\linewidth]{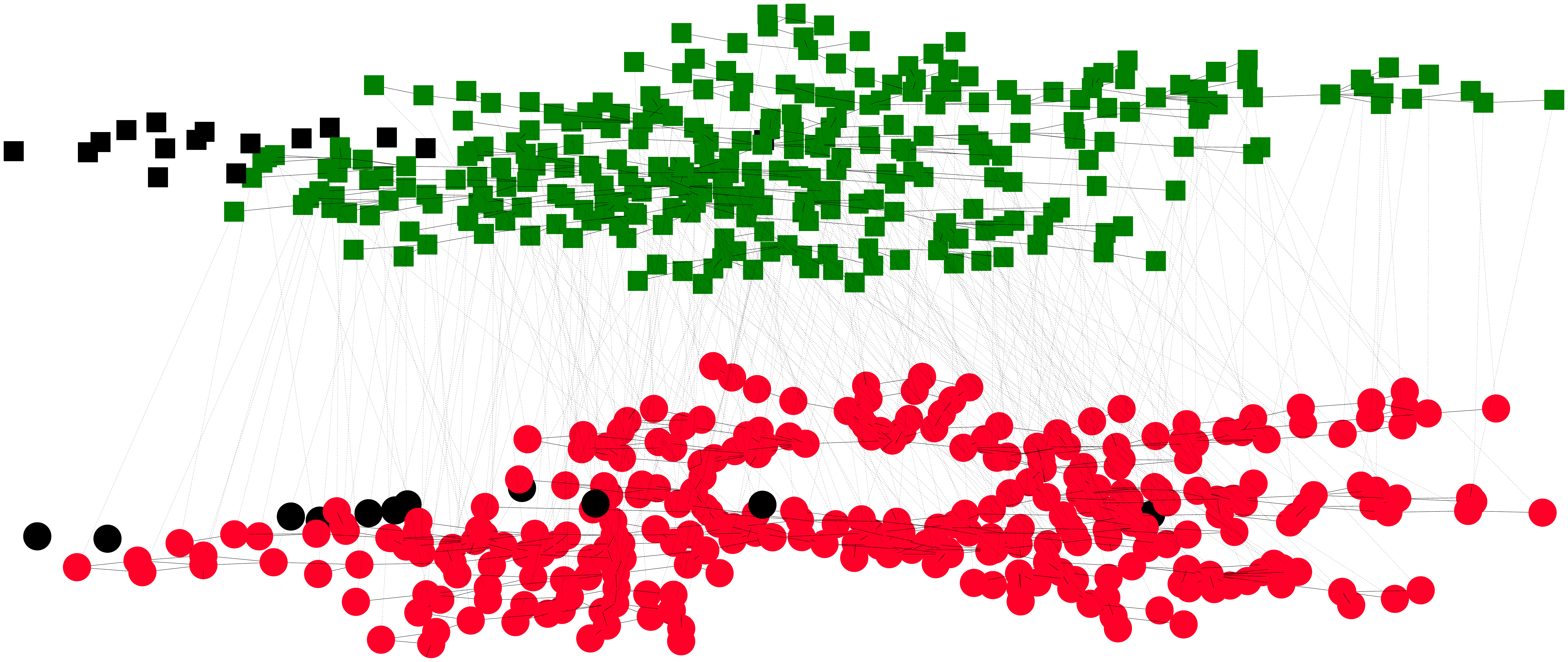}
	}
	\\ \vspace{\vsp}
	\subfloat[3A]{
		\centering
		\includegraphics[width=\wdt\linewidth]{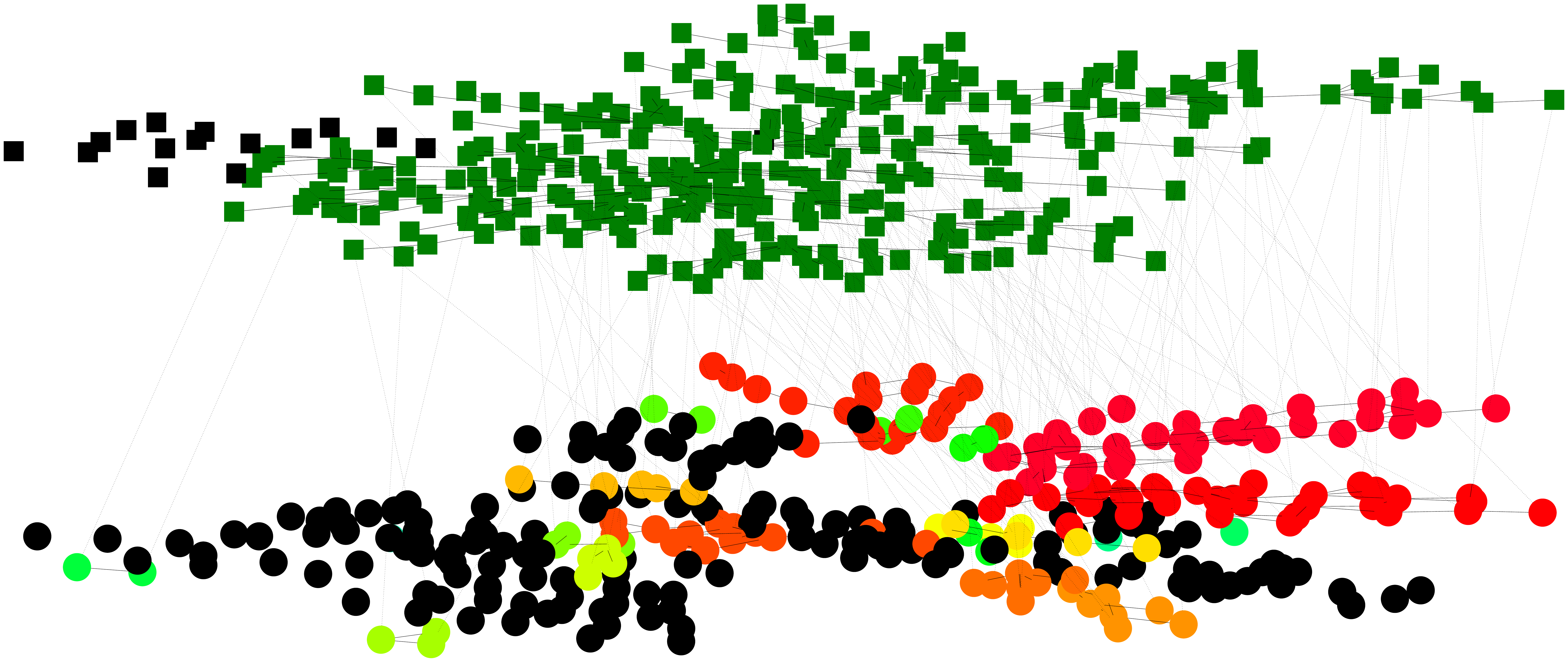}
	} 
	\subfloat[3C]{
		\centering
		\includegraphics[width=\wdt\linewidth]{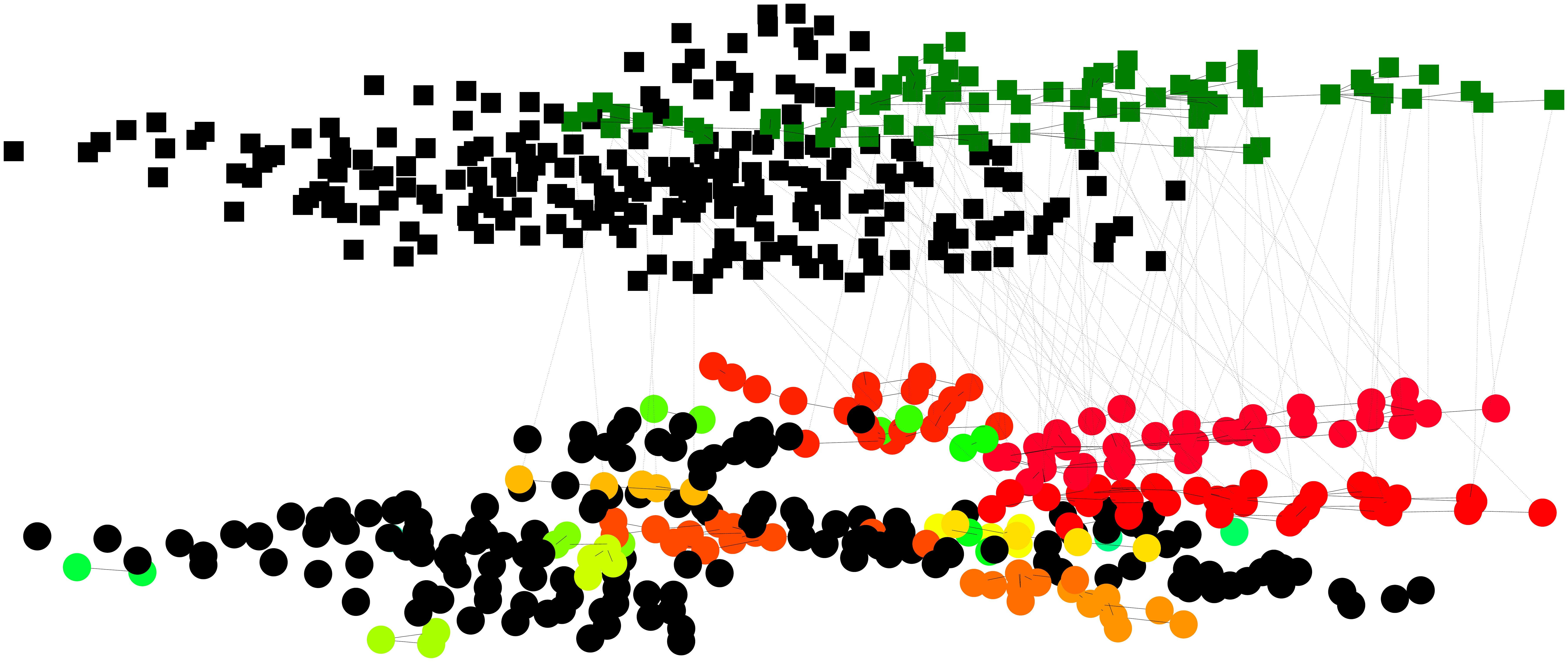}
	}
	\\ \vspace{\vsp}
	\subfloat[4A]{
		\centering
		\includegraphics[width=\wdt\linewidth]{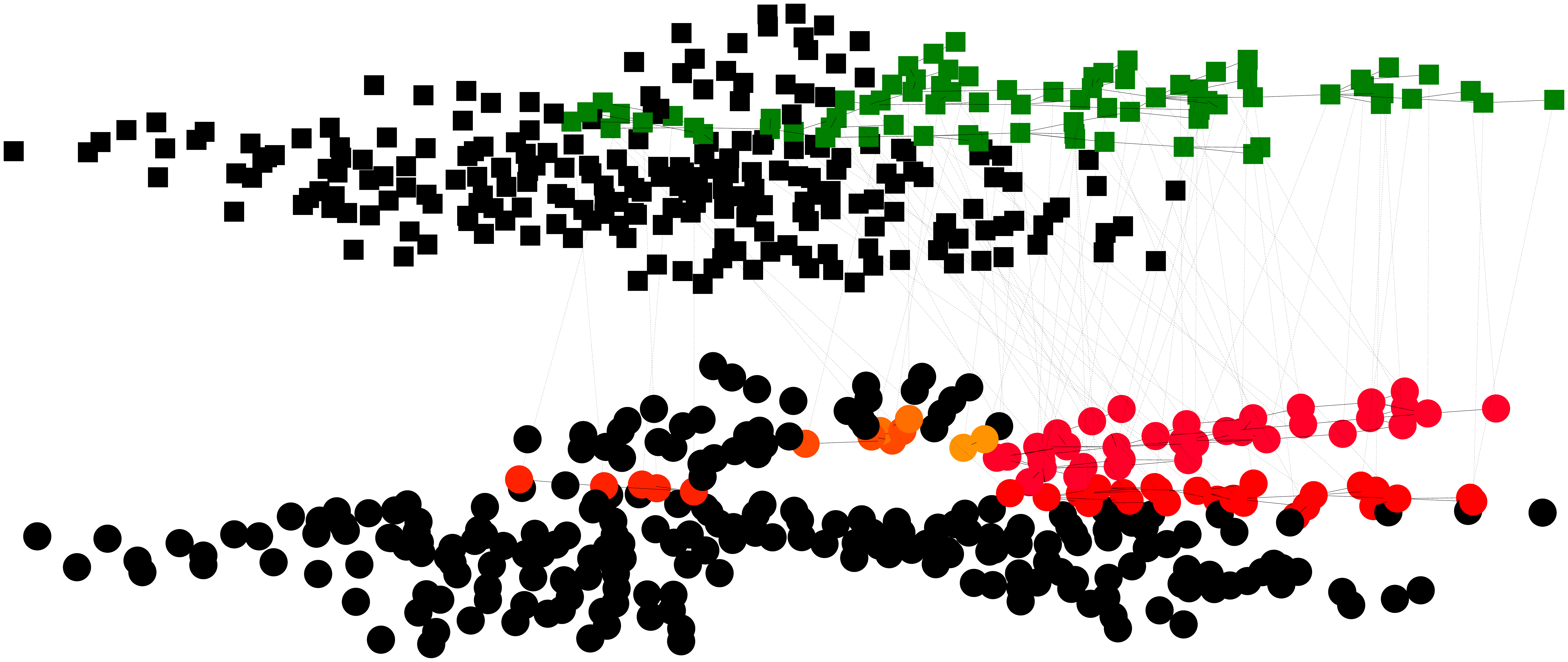}
	} 
	\subfloat[4C]{
		\centering
		\includegraphics[width=\wdt\linewidth]{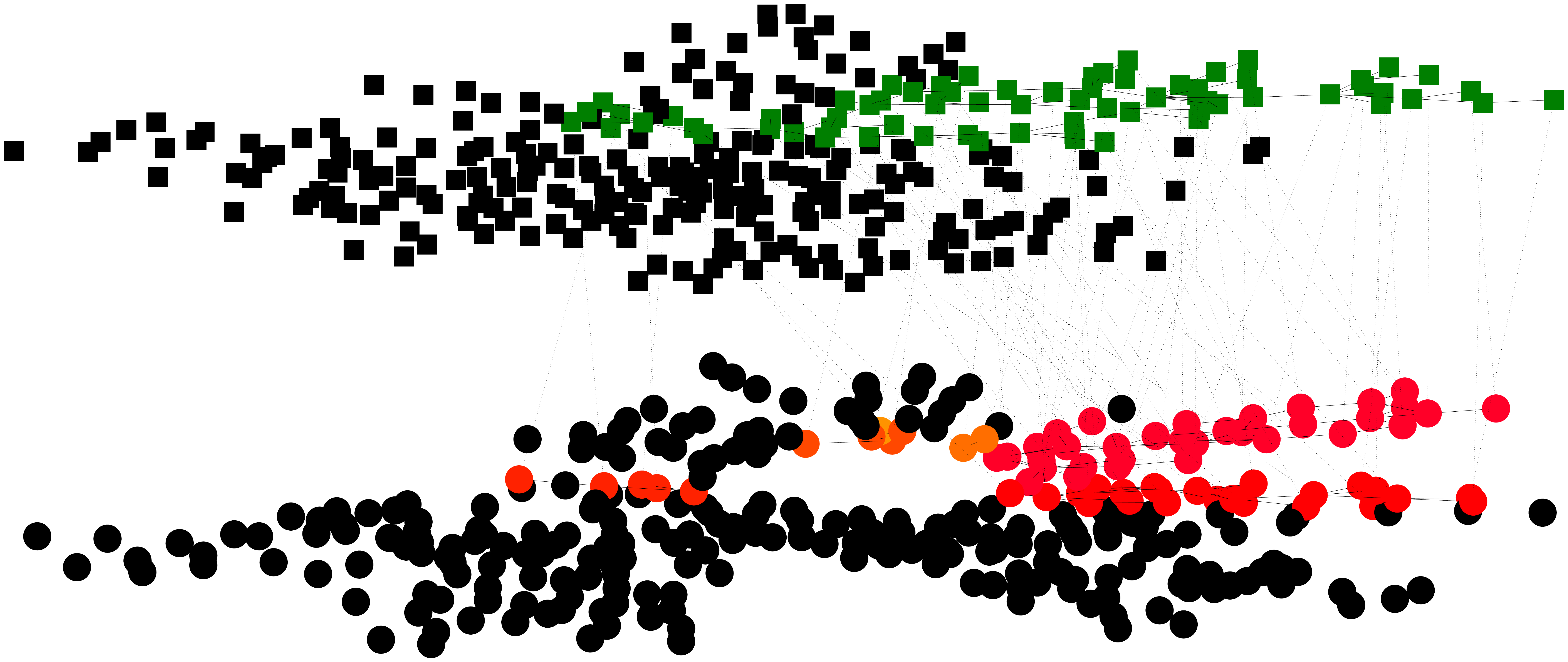}
	}
	\\ \vspace{\vsp}
	\subfloat[5A]{
		\centering
		\includegraphics[width=\wdt\linewidth]{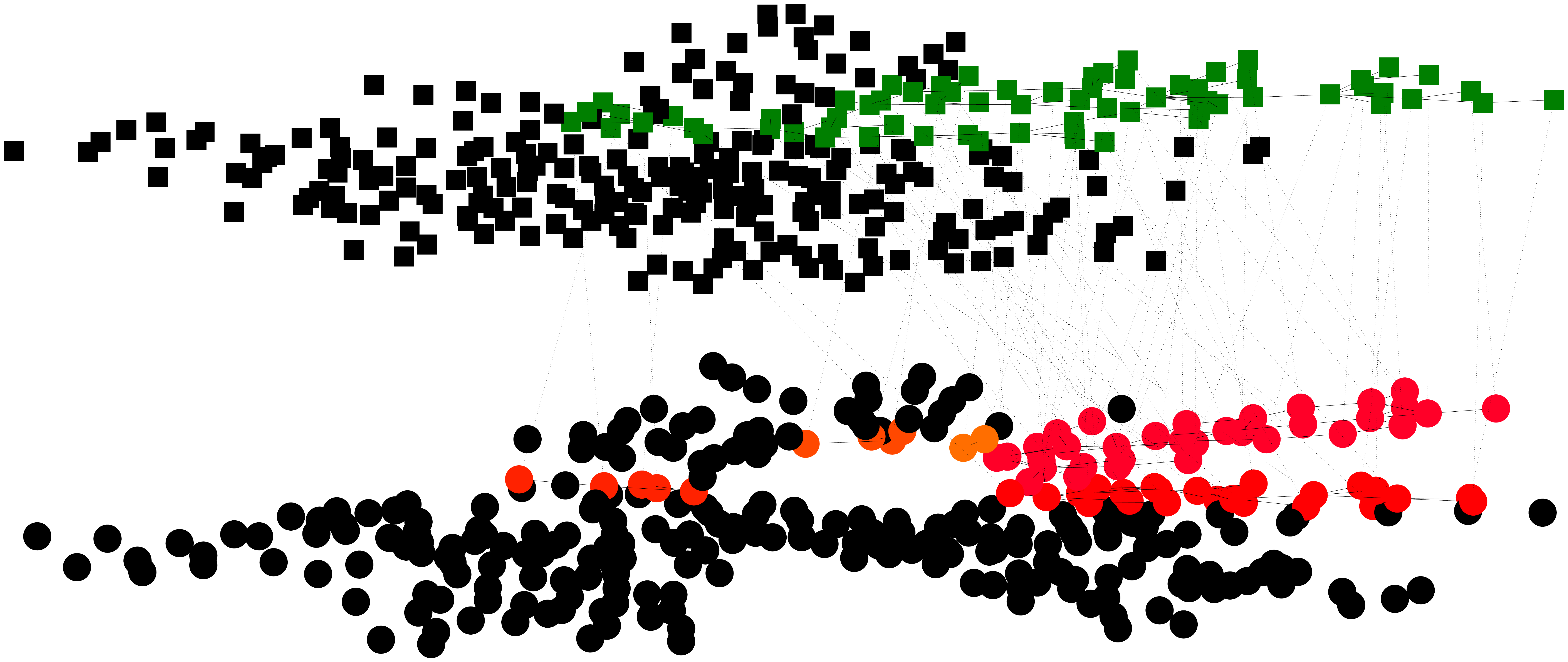}
	} 
	\subfloat[5C]{
		\centering
		\includegraphics[width=\wdt\linewidth]{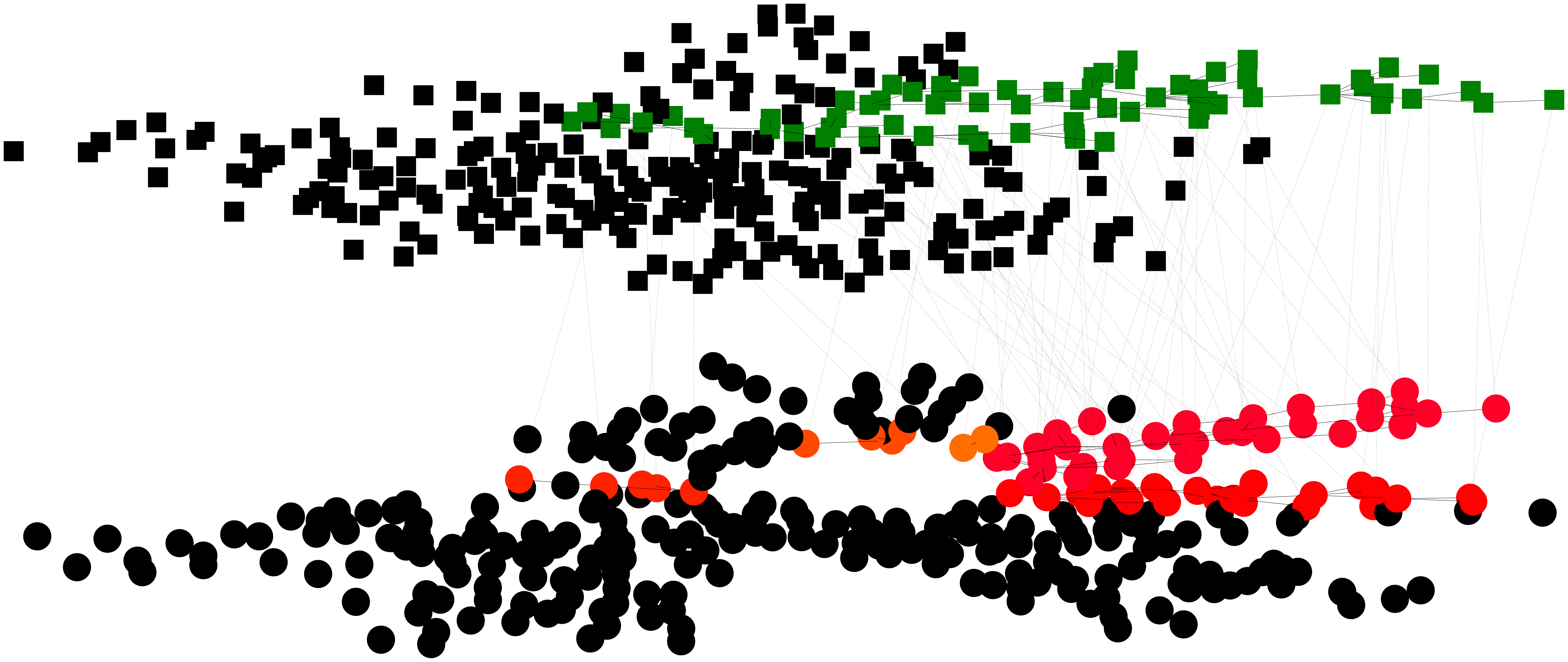}
	}
	\\ \vspace{\vsp}
	\caption{Different iterations of the proposed model for failure propagation in the IEEE 300-bus test system for a random attack. Note that the failure triggered in power system (b) and propagates in power system which causes three additional failures (c)bus which demonstrates that a failure can propagate non-locally due to the power system's physics.}
	\label{fig:it300}
\end{figure*}

Figs.~\ref{fig:it118} and \ref{fig:it300} visualize different phases of failure propagation after random attacks for IEEE 118- and 300-bus test cases, respectively. Comparing Fig\ref{fig:it300}(b) and Fig\ref{fig:it300}(c) demonstrate that the failures propagate nonlocally in power systems.


\section{Conclusion}
To better understand the dynamics of cascading failure in cyber-physical power systems, we proposed an asynchronous algorithm to model failure propagation in cyber-physical power systems. In contrast to most previous works that employ DC power flow model for cascading failure analysis, we implement the full AC power flow equation in our model to accurately address the physics of power systems. Moreover, to make the proposed approach more general, the interdependency between power and cyber layers is taken into consideration in the proposed model. We illustrated via simulations the failure propagation in cyber-physical power systems. We showed that the failures in power systems propagate globally, i.e., the consecutive failures can be far away from each other, rather than locally. We also show that taking cyber layer impacts into account can expedite failure propagation process, as it was expected. The outcome of this paper can be used to analyze the resiliency of cyber-physical systems against failure propagation. Furthermore, the proposed model can shed light on identifying critical components and analyzing contingency analysis in cyber-physical systems. This work is under study to model the interaction between power and cyber layers using graph network theory metrics. More specifically, dependability and importance of a specific node in one layer to the corresponding node in another layer will be incorporated in the proposed model to more accurately simulate the interaction between layers in cyber-physical power systems.     

\bibliographystyle{IEEEtran}
\bibliography{sgre2022}

\end{document}